\documentclass[aps,prl,twocolumn,showpacs]{revtex4}

\newcommand{\nc}{\newcommand}
\nc{\be}{\begin{equation}} \nc{\ee}{\end{equation}}
\nc{\bea}{\begin{eqnarray}} \nc{\eea}{\end{eqnarray}}
\nc{\bi}[1]{\bibitem{#1}}

\def\d{\delta}

\begin{document}


\leftline{\hspace{5.4in} UFIFT-QG-07-01}

\vskip 0.2in

\title{A Generic Test of Modified Gravity Models which Emulate Dark Matter}

\author{E. O. Kahya }
\email[]{kahya@phys.ufl.edu} \affiliation{Department of Physics,
University of Florida,
             Gainesville, FL 32611, USA}

\author{R. P. Woodard}
\email[]{woodard@phys.ufl.edu} \affiliation{Department of Physics,
University of Florida,
             Gainesville, FL 32611, USA}

\date{\today}

\begin{abstract}

We propose a generic test for models in which gravity is modified to
do away with dark matter. These models tend to have gravitons couple 
to a different metric than ordinary matter. A strong test of such 
models comes from comparing the arrival time of the gravitational 
wave pulse from a cosmological event such as a supernova with the 
arrival times of the associated pulses of neutrinos and photons. For 
SN 1987a we show that the gravity wave would have arrived 5.3 days 
after the neutrino pulse.

\end{abstract}

\pacs{04.80.Cc,95.85.Sz,95.85.Pw,95.35.+d}

\maketitle
%
%

\noindent \textbf{1. Introduction} \vskip 0.1in

As early as 1933 Zwicky was able to infer the need for dark matter 
by applying the virial theorem to observations of the Coma Cluster 
\cite{Zwi}. In the 1970's Rubin, Ford and Thonnard accumulated 
independent evidence from the rotation curves of spiral galaxies
\cite{RTF,RFT1,RFT2}. And the 1990's saw weak lensing used to probe
dark matter in galactic clusters \cite{TVW,FKSW,SKFBW,CLKHG,Mellier,WTKAB}.

It will be seen that the evidence for dark matter has so far been entirely 
restricted to gravity. That is, one infers the Einstein tensor from 
measurements of cosmic motions, or from lensing, and then compares the 
result with the observed stress-energy in stars and gas,
\begin{equation}
\Bigl( R_{\mu\nu} - \frac12 g_{\mu\nu} R\Bigr)_{\rm inf} = 
\frac{8\pi G}{c^4} \Bigl( T_{\mu\nu}\Bigr)_{\rm obs} \; . \label{Ein}
\end{equation}
There is indeed a disagreement between the two sides of the equation,
but it is not clear that this signifies the need for dark matter on the 
right of (\ref{Ein}) rather than a modification of gravity on the left. 
Although there are plausible dark matter candidates, none of them has 
so far been detected in a terrestrial laboratory \cite{CDMS1,CDMS2,ADMX}.

Certain regularities in cosmic structures suggest modified gravity.
One of these is the Tully-Fisher relation, which states that the
luminosity of a spiral galaxy is proportional to the fourth power 
of the peak velocity in its rotation curve \cite{TF}. If luminous 
matter is insignificant compared to dark matter, why should such a
relation exist? Another regularity is Milgrom's Law, which states
that the need for dark matter occurs at gravitational accelerations
of $a_0 \simeq 10^{-10}~{\rm m/s}^2$ \cite{KT}. This has been observed 
in cosmic structures which vary in size by six orders of magnitude
\cite{SM}.

A modification of Newtonian gravity which explains these regularities 
was proposed by Milgrom in 1983 \cite{Milg}. His model, Modified 
Newtonian Dynamics (MOND), was soon given a Lagrangian formulation in 
which conservation of energy, 3-momentum and angular momentum are 
manifest \cite{BM}. However, there was for years no successful 
relativistic generalization which could be employed to study 
cosmological evolution. Even in the context of static, spherically 
symmetric geometries, 
\begin{equation}
ds^2 \equiv -B(r) c^2 dt^2 + A(r) dr^2 + r^2 d\Omega^2 \; , \label{rgeom}
\end{equation}
the early formulation of MOND fixed only $B(r)$, not $A(r)$. It was 
therefore incapable of making definitive predictions about gravitational 
lensing.

A relativistic extension of MOND has recently been proposed by Bekenstein
\cite{Bek}. This model is known as TeVeS for ``Tensor-Vector-Scalar.''
In addition to reproducing the MOND force law at low accelerations,
TeVeS has acceptable post Newtonian parameters, and it gives a
plausible amount of gravitational lensing \cite{Bek}. When TeVeS is
used in place of general relativity + dark matter to study cosmological
evolution, the results are in better agreement with data than many
thought possible \cite{Skordis,AFZ,Dodelson,Bourliot}.

What concerns us here is the curious property of TeVeS that small amplitude
gravity waves are governed by general relativity through the metric 
$g_{\mu\nu}$, whereas matter couples to a ``disformally transformed'' metric 
which involves the vector and scalar fields,
\begin{equation}
\widetilde{g}_{\mu\nu} = e^{-2\phi} (g_{\mu\nu} + A_{\mu} A_{\nu}) 
- e^{2\phi} A_{\mu} A_{\nu} \; .
\end{equation}
The Scalar-Vector-Tensor gravity (SVTG) theory proposed by Moffat also
has different metrics for matter and small amplitude gravity waves
\cite{JWM,BrM}. The appearance of this feature in two very different
models is the result of trying to reconcile solar system tests with modified 
gravity at ultra-low accelerations. Solar system tests strongly predispose
the Lagrangian to possess an Einstein-Hilbert term \cite{Will}. On the other 
hand, failed attempts to generalize MOND \cite{SW1} have led to a theorem 
that one cannot get sufficient weak lensing from a stable, covariant and
purely metric theory which reproduces the Tully-Fisher relation without 
dark matter \cite{SW2}. Hence the MOND force must be carried by some other 
field, and it is a combination of this other field and the metric which 
determines the geodesics for ordinary matter. However, the dynamics of 
small amplitude gravity waves are still set by the linearized Einstein 
equation. This simple observation makes for a sensitive and generic test.

We define a {\it Dark Matter Emulator} as any modified gravity theory for 
which:
\begin{enumerate}
\item{Ordinary matter couples to the metric $\widetilde{g}_{\mu\nu}$
that would be produced by general relativity + dark matter; and}
\item{Small amplitude gravity waves couple to the metric $g_{\mu\nu}$ 
produced by general relativity without dark matter.}
\end{enumerate}
Now consider a cosmic event such as a supernova which emits simultaneous
pulses of gravity waves and either neutrinos or photons. If physics is 
described by a dark matter emulator then the pulse of gravity waves will 
reach us on a lightlike geodesic of $g_{\mu\nu}$, whereas neutrinos and
photons travel along a lightlike geodesic of $\widetilde{g}_{\mu\nu}$. 
If significant propagation occurs over regions that would be dark matter
dominated in general relativity then there will be a measurable lag
between arrival times. 

The magnitude of the expected time lag is so large that great precision 
is not needed for either $g_{\mu\nu}$ or $\widetilde{g}_{\mu\nu}$. We
demonstrate this in section 2 by computing the time lag for SN 1987a
using simple models for the distribution of luminous and dark matter.
Section 3 discusses the prospects for observation. Our conclusions 
comprise section 4.

\vskip 0.3in

\noindent \textbf{2. Time Lag} \vskip 0.1in

We model the luminous matter of our galaxy as a spherical bulge of $M \simeq 
10^{41}~{\rm kg}$. For this source general relativity gives a metric of
the form (\ref{rgeom}), with the functions $B$ and $A$ expressed in
terms of twice the Newtonian potential $\epsilon \equiv (2 G M)/(c^2 r)$,
\begin{equation}
g_{\mu\nu} \Longrightarrow B(r) = 1 - \epsilon \;\; , \; \;
A(r) = \frac1{1 - \epsilon} \simeq 1 + \epsilon \; .
\end{equation}
At the Sun's radius of $r_S \simeq 8.0~{\rm kpc}$ we have $\epsilon \simeq
6\times 10^{-7}$, so it is valid to neglect terms of order $\epsilon^2$.

To determine $\widetilde{g}_{\mu\nu}$ we assume the galaxy is surrounded 
by an isothermal halo which makes the asymptotic rotation speed $v_* = 
(a_0 G M)^{\frac14}$. This gives rise to a second small parameter,
$\epsilon_* \equiv 2 v_*^2/c^2 \simeq 6 \times 10^{-7}$. For this source
general relativity gives a metric of the form (\ref{rgeom}) with,
\begin{equation}
\widetilde{g}_{\mu\nu} \Longrightarrow \widetilde{B}(r) \simeq 
1 - \epsilon + \epsilon_* \ln\Bigl(\frac{r}{r_S}\Bigr) \; \; , \; \;
\widetilde{A}(r) \simeq 1 + \epsilon + \epsilon_* \; . \label{Btilde}
\end{equation}
We have chosen the integration constant in the order $\epsilon_*$ 
contribution to $\widetilde{B}(r)$ so that it vanishes at $r = r_S$. In 
section 4 we discuss the effect of other choices.

It is most efficient to work this problem in Cartesian coordinates for
which the nonzero components of the two metrics are,
\begin{eqnarray}
g_{00} = -1 + \epsilon &,& g_{ij} \simeq \delta^{ij} + \epsilon \widehat{r}^i 
\widehat{r}^j \; , \\
\widetilde{g}_{00} \simeq -1 + \epsilon + \epsilon_* \ln\Bigl(\frac{r}{r_S}
\Bigr) &,& \widetilde{g}_{ij} \simeq \delta^{ij} + (\epsilon + \epsilon_*) 
\widehat{r}^i \widehat{r}^j . \quad
\end{eqnarray}
Here the radial unit vector is $\widehat{r}^i \equiv x^i/r$, the radius
is $r = \Vert \vec{x} \Vert$ and we have neglected terms which are 
higher than linear in either $\epsilon$ or $\epsilon_*$. 

In each metric we must construct the lightlike geodesic from a point 
$\vec{x}_L$ in the Large Magellanic Cloud to the Sun's position $\vec{x}_S$. 
For $g_{\mu\nu}$ this means solving the second order equation,
\begin{equation}
\ddot{\chi}^{\mu}(\tau) + \Gamma^{\mu}_{~\rho\sigma}\Bigl(\chi(\tau)\Bigr)
\dot{\chi}^{\rho}(\tau) \dot{\chi}^{\sigma}(\tau) = 0 \; . \label{geqn}
\end{equation}
The general solution to (\ref{geqn}) depends upon eight integration constants 
which can be taken as the initial position $\chi^{\mu}(0)$ and velocity 
$\dot{\chi}^{\mu}(0)$. Because $g_{\mu\nu}$ is static we may as well start 
the geodesic at $t=0$,
\begin{equation}
\chi^0(0) = 0 \qquad , \qquad \chi^i(0) = x^i_L \; .
\end{equation}
We fix the spatial components of the initial velocities by requiring the
geodesic to reach the Sun at $\tau = 1$,
\begin{equation}
\chi^i(1) = x^i_S \qquad \Longleftrightarrow \qquad \dot{\chi}^i(0) \; .
\label{space}
\end{equation}
The initial temporal velocity is fixed by requiring the geodesic to be
lightlike,
\begin{equation}
g_{\mu\nu}(0,\vec{x}_L) \dot{\chi}^{\mu}(0) \dot{\chi}^{\nu}(0) = 0 \qquad
\Longleftrightarrow \qquad \dot{\chi}^0(0) \; . \label{light}
\end{equation}
The geodesic $\widetilde{\chi}^{\mu}(\tau)$ is determined the same way, 
but with the substitutions of $\widetilde{\Gamma}^{\mu}_{~\rho\sigma}$ for 
$\Gamma^{\mu}_{~\rho\sigma}$ in (\ref{geqn}) and of $\widetilde{g}_{\mu\nu}$ 
for $g_{\mu\nu}$ in (\ref{light}). The time lag we seek is,
\begin{equation}
\Delta t \equiv \frac1{c} \Bigl(\widetilde{\chi}^0(1) - \chi^0(1)\Bigr) \; .
\label{Dt}
\end{equation}

Because $\epsilon$ and $\epsilon_*$ are so small we can use perturbation
theory. The $0^{\rm th}$ order (flat space) solution is the same for both 
geodesics,
\begin{equation}
\overline{\chi}^0(\tau) = \Delta x \, \tau \qquad , \qquad \overline{\chi}^i
= x^i_L + {\Delta x}^i \tau \; ,
\end{equation}
where $\Delta x^i \equiv x^i_S - x_L^i$. Because both $g_{\mu\nu}$ and 
$\widetilde{g}_{\mu\nu}$ have the same order $\epsilon$ terms, the order
$\epsilon$ corrections to $\chi^{\mu}$ cancel those to $\widetilde{\chi}^{\mu}$
in (\ref{Dt}) and we need only consider the corrections of order $\epsilon_*$. 
The nonzero components of the order $\epsilon_*$ connection are, 
\begin{equation}
\Delta \Gamma^0_{~0i} \!=\! \Delta \Gamma^i_{~00} \!=\! 
\frac{\epsilon_* \widehat{r}^i}{2r} \; , \; \Delta \Gamma^i_{~jk} \!=\!
\frac{\epsilon_* \widehat{r}^{i}}{r} \!\Bigl[\d^{jk} \!-\! \widehat{r}^{j}
\widehat{r}^{k} \Bigr] . \label{Christoffel}
\end{equation}
The $1^{\rm st}$ order corrected geodesic follows from integrating the
geodesic equation,
\begin{eqnarray}
\lefteqn{\Delta \chi^{\mu}(\tau) \equiv \widetilde{\chi}^{\mu}(\tau) -
\chi^{\mu}(\tau) \; ,} \\
& & = \Delta \dot{\chi}^{\mu}(0) \!-\! \int_0^{\tau} \!\!\!\! d\tau' \!\! 
\int_0^{\tau'} \!\!\!\! d\tau'' \, \Delta \Gamma^{\mu}_{~\rho\sigma}\Bigl(
\overline{\chi}(\tau'')\Bigr) \dot{\overline{\chi}}^{\rho} 
\dot{\overline{\chi}}^{\sigma} \; . \qquad
\end{eqnarray}
The spatial boundary conditions (\ref{space}) imply,
\begin{equation}
\Delta \dot{\chi}^i(0) = \int_0^1 \!\!\! d\tau \, (1-\tau) 
\Delta \Gamma^i_{~\rho\sigma}\Bigl(\overline{\chi}(\tau)\Bigr) 
\dot{\overline{\chi}}^{\rho} \dot{\overline{\chi}}^{\sigma} \; .
\end{equation}
The lightlike condition (\ref{light}) fixes the final integration constant,
\begin{equation}
\Delta \dot{\chi}^0(0) = \frac{\Delta x^i \Delta \dot{\chi}^i(0)}{\Delta x}
+ \frac{\epsilon_* \Delta x}{2} \ln\Bigl(\frac{r_L}{r_S}\Bigr) + 
\frac{\epsilon_* (\widehat{r} \cdot \Delta \vec{x})^2}{2 \Delta x} \; .
\end{equation}
The parameter integrations are straightforward when expressed in terms 
of the dimensionless constants,
\begin{equation}
\alpha \equiv \frac{\vec{x}_L \cdot \Delta \vec{x}}{\Delta x^2} 
\qquad {\rm and} \qquad \beta \equiv \frac{r_L^2}{\Delta x^2} \; .
\end{equation}
Our result for the first order time lag is,
\begin{eqnarray}
\lefteqn{\Delta t = \frac{\epsilon_* \Delta x}{c} \Biggl[1 +
\frac{\alpha}2 \ln\Bigl(\frac{r_L}{r_S}\Bigr) } \nonumber \\
& & \hspace{2cm} - \sqrt{\beta - \alpha^2} \, \tan^{-1}\!\!\left(
\frac{\sqrt{\beta - \alpha^2}}{\beta + \alpha}\right) \Biggr] . 
\qquad \label{fineq} 
\end{eqnarray}

Within the context of our simple model for the matter distributions,
equation (\ref{fineq}) is valid for any $\vec{x}_L$. The specific
values appropriate to SN 1987a are \cite{SN1987a},
\begin{eqnarray} 
\lefteqn{r_S \simeq 8.0~{\rm kpc} \quad , \quad r_L \simeq 50.9~{\rm kpc} 
\quad , \quad \Delta x \simeq 51.4~{\rm kpc} \; ,} \nonumber \\
& & \hspace{1.5cm} \alpha \simeq -.9775 \qquad {\rm and} \qquad 
\beta \simeq .9793 \; . \qquad 
\end{eqnarray}
Substituting these values in (\ref{fineq}) gives,
\begin{equation}
\Delta t \Bigl\vert_{\scriptscriptstyle {\rm SN\ 1987a}} \simeq 
36.7~{\rm days} \, \Bigl[ -.144\Bigr] \simeq -5.3~{\rm days} \; .
\end{equation}
Therefore simultaneous pulses of neutrinos and/or photons would 
have arrived 5.3 days before the gravity waves if physics were described
by a dark matter emulator.

The sign of (\ref{fineq}) is mostly set by the difference of the two 
gravitational potentials, $\widetilde{B}(r) - B(r) \simeq \epsilon_* 
\ln(r/r_S)$. The lag is negative because ordinary matter is at higher
potential than gravity waves. The square bracketed term in (\ref{fineq}) 
is typically of order one. (It is $-.144$ for SN 1987a.) Hence the rough 
magnitude of the time lag is $\epsilon_* \simeq 6 \times 10^{-7}$ times 
the flat space propagation time. This works out to be $36.7~{\rm days}$ 
for SN 1987a. One could therefore anticipate a lag in the range of days 
without the detailed computation. \vskip 0.3in

\noindent \textbf{3. Observational Prospects} \vskip 0.1in

The test we have proposed requires detecting gravity waves and either
neutrinos or photons that were emitted simultaneously from some
cosmic event separated from us by an extensive region in which the
gravitational influence of dark matter predominates over that of
luminous matter. If physics is described by general relativity + dark
matter then the arrival times will be simultaneous. If physics 
is instead described by a dark matter emulator then the photon and
neutrino pulses will generally arrive some days before the gravity 
wave signal.

One only expects a few supernovae per century in our galaxy. Of course 
the light signal from a supernova can be observed even from distant 
galaxies, but this is much further than we have any hope of detecting 
neutrinos or gravity waves. The optical signal from a supernova will 
typically lag the neutrino pulse by several hours --- as was the case 
for SN 1987a \cite{SN1987a} --- because photons must traverse the 
optically dense stellar environment. However, the magnitude of the 
expected delay for the gravity wave signal is so great that this should 
not matter except for nearby supernovae.

The amount of gravitational radiation from a supernova depends on
the oblateness of the progenitor. If the oblateness of SN 1987a was
in relation to that of the Sun then current gravitational wave 
detectors would probably not have seen anything \cite{Dimmel}. However, 
advanced LIGO would detect such a supernova out to $0.8~{\rm Mpc}$ 
\cite{Dimmel}. This includes the Andromeda galaxy, which doubles the
expected rate and also ensures that the signal passes through dark
matter dominated regions.

Neutrinos from SN 1987a were observed by the Kami\-o\-kan\-de-II 
\cite{Kam1,Kam2} and Irvine-Michigan-Brookhaven \cite{Bion,Brat} detectors. 
Of course the state of the art has improved since then, although effective 
coverage is still limited to our own galaxy and its satellites 
\cite{BFV,Vogel}. There is now a Supernova Early Warning System 
\cite{SNEWS} comprised of Super-Kamiokande \cite{SNO1}, the Large Volume 
Detector \cite{LVD}, the Sudbury Neutrino Observatory \cite{SNO2} --- 
which is no longer operating but may be succeeded by SNO+ \cite{SNO+} --- 
and the IceCube Detector \cite{Ice}. \vskip 0.3in

\noindent \textbf{4. Discussion} \vskip 0.1in

Our largest uncertainty is modeling the radius at which the density of
dark matter reaches its asymptotic form of $\rho(r) \simeq v_*^2/(4\pi 
G r^2)$. We chose that radius to be $r_S$ in $(\ref{Btilde})$ because
this is near where the Newtonian acceleration reaches the value of $a_0 
\simeq 10^{-10}~{\rm m/s}^2$ at which Milgrom's Law implies the need for
dark matter \cite{KT}. Had we chosen some other radius, $r_O$, the square 
bracketed term in (\ref{fineq}) would change to
\begin{equation}
\Bigl[\qquad\Bigr] \longrightarrow \Bigl[\qquad\Bigr] -\frac12 
\ln\Bigl(\frac{r_S}{r_O}\Bigr) \; .
\end{equation}
One would expect this to make the time lag more negative because it is 
difficult to imagine having dark matter ``turn on'' much beyond $r_S$.

If neutrinos have mass then a neutrino of energy $E$ would propagate
with velocity,
\begin{equation}
v \;=\; c \sqrt{1 - \frac{m^2 c^4}{E^2}} \label{vel}. 
\end{equation}
Taking the average detected supernova neutrino energy to be $E \simeq 
10~{\rm MeV}$ \cite{Yuksel}, and the (electron) neutrino mass to be its 
current upper limit of $m \simeq 2~{\rm eV}$ \cite{PDB}, the velocity 
would be changed by $\Delta v \simeq -2 \times 10^{-14} c$. Over a 
distance of $50~{\rm kpc}$ this would delay the neutrino by $.1~{\rm s}$, 
which is negligible on the scale of days expected for the primary effect.

The generality of our test derives from avoiding the details of specific
models and concentrating instead upon what they do, which is to emulate
dark matter. Rotation curves and lensing are too well measured for any 
viable model not to somehow reproduce the metric $\widetilde{g}_{\mu\nu}$ 
which couples to ordinary matter. We have explained the theoretical
reasons why gravity waves are predisposed to couple to the original metric 
$g_{\mu\nu}$ \cite{Will,SW2}. Both of the existing models do have this 
property \cite{Bek,JWM}, but as we cannot prove it to be unavoidable, it
is safest to comment that our results are limited to dark matter 
emulators. \vskip 0.3in

\noindent \textbf{Acknowledgements} \vskip 0.1in

We are grateful for conversations and correspondence with L. Blanchet,
H. Ray, C. Skordis and B. Whiting. This work was partially supported by 
NSF grant PHY-0244714 and by the Institute for Fundamental Theory at the 
University of Florida.

\end{document}